\documentclass[preprint,showpacs,preprintnumbers,amsmath,amssymb,showkeys]{revtex4}
\def\texpsfig#1#2#3{\vbox{\kern #3\hbox{\includegraphics{#1}\kern #2}}\typeout{(#1)}}

\input epsf   % load the epsf library

\usepackage{graphicx}% Include figure files
\usepackage{dcolumn}% Align table columns on decimal point
\usepackage{bm}% bold math
\begin{document}

\title{Sine-Gordon Expectation Values of Exponential Fields With Variational Perturbation Theory}

\author{Wen-Fa Lu}
   \email{wenfalu@sjtu.edu.cn}
\affiliation{ \ Institute of Theoretical Physics, Department of
Physics, Shanghai Jiao Tong University, Shanghai 200030, the
People's Republic of China \ (home institute)} \affiliation{ \ The
Abdus Salam International Centre for Theoretical Physics, Strada
Costiera, II 34014, Trieste, Italy}
\date{\today}

\begin{abstract}
In this letter, expectation values of exponential fields in the
2-dimensional Euclidean sine-Gordon field theory are calculated
with variational perturbation approach up to the second order. Our
numerical analysis indicates that for not large values of the
exponential-field parameter $a$, our results agree very well with
the exact formula conjectured by Lukyanov and Zamolodchikov in
Nucl. Phys. B 493, 571 (1997).
\end{abstract}

 \maketitle

\section{Introduction}
\label{1}

This letter briefly reports our investigation on vacuum
expectation value (VEV) of the exponential field
$e^{ia\phi(\vec{r})}$ in the sine-Gordon (sG) field theory, $G_a$,
with variational perturbation theory (VPT). Here, $a$ is a
parameter, and $\phi(\vec{r})$ is the field operator at the
2-dimensional Euclidean space (2DES) point $\vec{r}=(x,\tau)$
($\tau$ is the Euclidean time).

In 1997, starting from the exact expressions for the three special
cases: the coupling $\beta\to 0$ (semi-classical limit),
$\beta={\frac {1}{2}}$ and $a=\beta$, Lukyanov and Zamolodchikov
guessed an exact formula for the VEV of $e^{ia\phi(\vec{r})}$ in
the sG field theory at any $\beta^2<1$ and $|Re(a)|<1/(2\beta)$
\cite{1}. Then, defining ``fully connected" one-point functions,
$\sigma_{2n}$ ($n$ is any natural number), from the VEV's of
even-power fields $\phi^{2n}$, they showed that $\sigma_{2}$ and
$\sigma_{4}$ from the above-mentioned exact formula agree with
those from perturbation theory for the sG field theory up to
$\beta^4$ and that $\sigma_{2}$ agrees with the corresponding
one-point function from perturbation theory up to $g$ the coupling
in the massive Thirring model, which is the fermion version of the
sG field theory \cite{2}. Furthermore, based on the reflection
relations with Liouville reflection amplitude \cite{3}, some extra
arguments for supporting the conjectured exact formula were
presented by their collaborators with them in the subsequent
papers \cite{4}. Slightly later, in 2000, checks from perturbation
theories in both an angular and a radial quantization approaches
for the massive Thirring model indicated that the perturbation
result up to $g$ exactly coincides with the corresponding result
obtained by expanding the exact formula according to the coupling
$g$ \cite{5}. Additionally, a numerical study for the model in
finite volume also provides evidence for the the conjectured exact
formula \cite{5a}. Thus, these investigations have given the
conjectured exact formula for the VEV of $e^{ia\phi(\vec{r})}$ a
complete check for the case of $\beta\to {\frac {1}{2}}$ ($g\to 0$
is equivalent to $\beta^2\to {\frac {1}{2}}$, from Eq.(8) in
Ref.~\cite{1}), and some indirect evidences for its validity.
Obviously, a direct check for the cases of $\beta^2\not={\frac
{1}{2}}$ (Ref.~\cite{1} has provided a partial check for the case
of $\beta\to 0$) is still needed.

In fact, VPT \cite{6} may provide such a check. It is some kind of
expansion theory similar to the perturbation theory. However,
because it properly ``grafts" the variational approach onto the
perturbation theory, the VPT produces non-perturbative results
which are valid for any coupling strength, including both weak and
strong couplings. Moreover, because the principle of minimal
sensitivity (PMS) \cite{7} is used to determine a man-made
parameter, the VPT is believed to be a convergent theory, and its
approximate accuracy can be systematically controlled and improved
to tend towards the exact \cite{6}. The primitive idea of VPT is
not a new one and can date back to 1955 at least \cite{8}. Now, it
has developed with many equivalent practical schemes, and has been
applied to quantum field theory (QFT), condensed matter physics,
statistical mechanics, chemical physics, and so on \cite{6}. In
this letter, based on the variational perturbation scheme in one
of our former joint papers \cite{9} (the scheme in Ref.~\cite{9}
was stemmed from the Okopinska's optimized expansion \cite{10},
and proposed by Stancu and Stevenson \cite{11}), we will develop a
variational perturbation scheme for the purpose of the present
letter and calculate approximately $G_a$ up to the second order.
No explicit divergences exist in the resultant expression owing to
the adoption of the Coleman's normal-ordering renormalization
prescription \cite{2,12,9}. One will see that our investigation
strongly support the conjectured exact formula.

Next section, we will develop the VPT to calculate the VEV of
$e^{ia\phi(\vec{r})}$ and give the VEV of $e^{ia\phi(\vec{r})}$ up
to the second order in VPT. In section III, we will report our
numerical results and make a numerical comparison between our
results and the exact ones. A brief conclusion will be made in
Sect. V.

\section{VPT for VEV's and Approximate $G_a$ up to the Second Order}
\label{2}

We consider the 2-dimensional Euclidean sG field theory with the
following Lagrangian density
\begin{equation}
{\cal L}_{sG}= {\frac {1}{2}}\nabla_{\vec{r}}\phi_{\vec{r}}
\nabla_{\vec{r}} \phi_{\vec{r}} - 2
\Omega\cos(\sqrt{8\pi}\beta\phi_{\vec{r}})\;.
\end{equation}
In this letter, the subscript $\vec{r}$ represents the coordinate
argument, for example, $\phi_{\vec{r}}\equiv \phi(\vec{r})$, and
$\nabla_{\vec{r}}$ is the gradient in the 2DES. The Lagrangian
density Eq.(1) is nothing but Eq.(5) in Ref.~\cite{1} if one makes
the transform $\phi\to {\frac {\phi}{\sqrt{8\pi}}}$ (hereafter, we
will use $e^{i\sqrt{8\pi}a\phi(\vec{r})}$ instead of
$e^{ia\phi(\vec{r})}$ and consequently the parameter $a$ and the
coupling $\beta$ in this letter are identical to those in
Ref.~\cite{1}, respectively). If taking $ \sqrt{8\pi}\beta \to
\beta$ and $2\Omega=m^2/\beta^2$ and adding the term $m^2/\beta^2$
in the Lagrangian density, one will get the Euclidean version of
the sG Lagrangian density which discussed in Ref.~\cite{9}. In
Eq.(1), the dimensionless $\beta$ is the coupling parameter and
$\Omega$ is another parameter with the dimension [lenth$]^{-2}$ in
the natural unit system. It is always viable to have $\beta\ge 0$
without loss of generality. The classical potential
$V(\phi_{\vec{r}})=-2 \Omega\cos(\sqrt{8\pi}\beta\phi_{\vec{r}})$
is invariant under the transform $\phi\to \phi+{\frac {2\pi
n}{\sqrt{8\pi}\beta}}$ with any integer $n$, and so the classical
vacua are infinitely degenerate. So do the quantum vacua of the sG
field theory according to Ref.\cite{9}. Here, as did in
Ref.\cite{1}, we choose to consider the symmetry vacuum with the
expectation value of the sG field operator $\phi_{\vec{r}}$
vanishing instead of those spontaneous symmetry broken vacua.

The VEV of the exponential field $e^{i\sqrt{8\pi}a\phi(\vec{r})}$
is defined as follows \cite{1}
\begin{equation}
G_a\equiv <e^{i \sqrt{8\pi}a \phi(0)}>\equiv {\frac {\int {\cal
D}\phi \exp\{i \sqrt{8\pi}a \phi(0)\} \exp\{-\int d^2 \vec{r}{\cal
L}_{sG}\}} {\int {\cal D}\phi \exp\{-\int d^2 \vec{r}{\cal
L}_{sG}\}}}  \;.
\end{equation}
For simplicity, the exponential field in Eq.(2) is taken at $r=0$.
It is evident that the numerator and denominator in the right hand
side of Eq.(2) can easily be got from the following sG generating
functional
\begin{equation}
Z[J]=\int {\cal D}\phi \exp\{-\int d^2 \vec{r}\; [{\cal
L}_{sG}-J_{\vec{r}}\phi_{\vec{r}}\; ]\}  \; ,
\end{equation}
by taking $J_{\vec{r}}=i \sqrt{8\pi}a \delta(\vec{r})$
($\delta(\vec{r})\equiv \delta(x)\delta(\tau)$) and $J=0$,
respectively. In Eq.(3), $J_{\vec{r}}$ is an external source at
$\vec{r}$. For renormalizing $G_a$, we will use its
normal-ordering form \cite{9,13}.

To perform a variational perturbation expansion on $G_a$, now we
modify $Z[J]$ in Eq.(3) by following Ref.~\cite{9} only without
shifting the field $\phi_{\vec{r}}$ (This is not necessary here
because we choose to consider the symmetrical vacuum as
aforementioned). That is, first introduce a parameter $\mu$ by
adding a vanishing term $\int d^2 \vec{r}{\frac
{1}{2}}\phi_{\vec{r}}(\mu^2-\mu^2)\phi_{\vec{r}}$ into the
exponent of the functional integral in Eq.(3), then rearrange the
exponent into a free-field part plus a new interacting part, and
finally insert a formal expansion factor $\epsilon$ before the
interacting part. Consequently, one has
\begin{eqnarray}
Z[J]\to Z[J;\epsilon]&=&\exp\{-\int d^2 \vec{r}\; [{\frac {1}{2}}
I_{(0)}(\mu^2)-{\frac {1}{2}}I_{(0)}({\cal M}^2)+{\frac
{1}{2}}{\cal M}^2 I_{(1)}({\cal M}^2)]\} \nonumber
\\ && \times \exp\{-\epsilon\int d^2 \vec{r}\; {\cal H}_I({\frac {\delta}{\delta J_{\vec{r}}}},\mu)\}
\exp\{{\frac {1}{2}}Jf^{-1}J\} \;
\end{eqnarray}
with
\begin{equation}
{\cal H}_I(\phi_{\vec{r}},\mu)=-{\frac
{1}{2}}\mu^2\phi_{\vec{r}}^2-2\Omega\cos(\sqrt{8\pi}\beta\phi_{\vec{r}})\exp\{4\pi\beta^2
I_{(1)}({\cal M}^2)\} \;.
\end{equation}
Here, ${\cal M}$ is a normal-ordering mass,
\begin{eqnarray*} I_{(n)}(Q^2)\equiv\Biggl \{
    \begin{array}{ll}
\int {\frac {d^2 \vec{p}}{(2\pi)^2}}
{\frac {1}{(p^2+Q^2)^n}} \;, & \ \ \ \ \ for \ \ n\not=0  \\
 \int {\frac {d^2 p}{(2\pi)^2}} \ln(p^2+Q^2)  \;, & \ \ \ \ \
                          for \ \  n=0
    \end{array}
\end{eqnarray*}
with $\vec{p}$ a Euclidean momentum in 2DES, and
$Jf^{-1}J\equiv\int d^2 \vec{r}' d^2 \vec{r}''
J_{\vec{r}'}f^{-1}_{\vec{r}' \vec{r}''}J_{\vec{r}''}$ with
\begin{equation}
f^{-1}_{\vec{r}' \vec{r}''}=\int {\frac {d^2 \vec{p}}{(2\pi)^2}}
{\frac {1}{p^2+\mu^2}} \; e^{i\vec{p}\cdot
(\vec{r}''-\vec{r}')}={\frac
{1}{2\pi}}K_0(\mu|\vec{r}''-\vec{r}'|) \;.
\end{equation}
In Eq.(6), $K_n(z)$ is the $n$th-order modified Bessel function of
the second kind. Note that Eq.(4) in the extrapolating case of
$\epsilon=1$ is only the normal-ordering expression of Eq.(3).

Expanding $\exp\{-\epsilon\int d^2 \vec{r}\; {\cal H}_I({\frac
{\delta}{\delta J_{\vec{r}}}},\mu)\}$ in Eq.(4) with Taylor series
of the exponential, one can write normal-ordered $G_a$ as
\begin{equation}
G_a=e^{4\pi a^2 I_{(1)}({\cal M}^2)}{\frac
{\bigl[\sum_{n=0}^\infty \epsilon^n {\frac {(-1)^{n}}{n!}} \int
\prod_{k=1}^n d^2 \vec{r}_k {\cal H}_I({\frac {\delta}{\delta
J_{\vec{r}_k}}},\mu)\exp\{{\frac
{1}{2}}Jf^{-1}J\}\bigl]_{J_{\vec{r}}=i \sqrt{8\pi}a
\delta(\vec{r})}}{\bigl[\sum_{n=0}^\infty \epsilon^n {\frac
{(-1)^{n}}{n!}} \int \prod_{k=1}^n d^2 \vec{r}_k {\cal H}_I({\frac
{\delta}{\delta J_{\vec{r}_k}}},\mu)
     \exp\{{\frac {1}{2}}Jf^{-1}J\}\bigl]_{J=0}}}\Biggl|_{\epsilon=1} \;.
\end{equation}
Thus, according to the formula 0.313 on page 14 in Ref.~\cite{14},
$G_a$ can be expanded as the following series of $\epsilon$
\begin{equation}
G_a=\bigl[G_a^{(0)}+\epsilon^1 G_a^{(1)}+\epsilon^2
G_a^{(2)}+\cdots +\epsilon^n G_a^{(n)} +\cdots\bigl]_{\epsilon=1}
\; .
\end{equation}

This series with $\epsilon=1$ is independent of the man-made
parameter $\mu$, but to get its closed form is beyond our ability.
Hence one can truncate it at any order of $\epsilon$ to
approximate it and then the truncated results will be dependent
upon $\mu$. This arbitrary parameter $\mu$ should be determined
according to the PMS \cite{7} as mentioned in the introduction.
That is, under the PMS, $\mu$ will be chosen from roots which make
the first (or second) derivative of the truncated result with
respect to $\mu$ vanish \cite{7,11,9}. Thus, $\mu$ will depend on
the truncated order. It is believed that it is this dependence
that makes the truncated result approach the exact $G_a$ order by
order \cite{6}. Thus the above procedure provides an approximate
method of calculating $G_a$ which can systematically control its
approximation accuracy. It is evident that this procedure has no
limits to the model coupling and is a non-perturbative method.

Executing the above procedure, we have truncated the series in
Eq.(8) at the second order of $\epsilon$, and the three
coefficients are
\begin{equation}
G_a^{(0)}=\exp\{4\pi a^2 I_{(1)}({\cal M}^2)\}\biggl[\exp\{{\frac
{1}{2}}Jf^{-1}J\}\biggl]_{J_{\vec{r}}=i \sqrt{8\pi}a
\delta(\vec{r})} \;,
\end{equation}
\begin{eqnarray}
G_a^{(1)}&=&-\exp\{4\pi a^2 I_{(1)}({\cal M}^2)\}\Biggl[\int d^2
\vec{r} {\cal H}_I({\frac {\delta}{\delta J_{r}}},\mu)
\exp\{{\frac {1}{2}}Jf^{-1}J\}\Biggl]_{J_{\vec{r}}=i \sqrt{8\pi}a
\delta(\vec{r})} \nonumber
\\ &&+G_a^{(0)}\Biggl[\int d^2 \vec{r} {\cal H}_I({\frac {\delta}{\delta
J_{r}}},\mu)\exp\{{\frac {1}{2}}Jf^{-1}J\}\Biggl]_{J_{\vec{r}}=0}
\;,
\end{eqnarray}
and
\begin{eqnarray}
G_a^{(2)}&=&{\frac {1}{2!}}e^{4\pi a^2 I_{(1)}({\cal
M}^2)}\Biggl[\int d^2 \vec{r}_1 d^2 \vec{r}_2 {\cal H}_I({\frac
{\delta}{\delta J_{\vec{r}_1}}},\mu){\cal H}_I({\frac
{\delta}{\delta J_{\vec{r}_2}}},\mu)\exp\{{\frac {1}{2}}Jf^{-1}J\}
\Biggl]_{J_{\vec{r}}=i \sqrt{8\pi}a \delta(\vec{r})}\nonumber \\
&&-{\frac {1}{2!}}G_a^{(0)}\Biggl[\int d^2 \vec{r}_1 d^2 \vec{r}_2
{\cal H}_I({\frac {\delta}{\delta J_{\vec{r}_1}}},\mu){\cal
H}_I({\frac {\delta}{\delta J_{\vec{r}_2}}},\mu)\exp\{{\frac
{1}{2}}Jf^{-1}J\} \Biggl]_{J_{\vec{r}}=0} \nonumber \\
&&+G_a^{(1)}\Biggl[\int d^2 \vec{r} {\cal H}_I({\frac
{\delta}{\delta J_{r}}},\mu) \exp\{{\frac
{1}{2}}Jf^{-1}J\}\Biggl]_{J_{\vec{r}}=0}\;,
\end{eqnarray}
respectively.

Performing carefully operations in Eqs.(9),(10) and (11), we
obtain the following expression of $G_a$ approximated up to the
second order of $\epsilon$, $G_a^{II}=[G_a^{(0)}+ \epsilon
G_a^{(1)}+ \epsilon^2 G_a^{(2)}]_{\epsilon=1}$,
\begin{eqnarray}
G_a^{II}&=& ({\frac {\mu^2}{{\cal M}^2}})^{a^2}(1-2 a^2
K_{02})+4\pi {\frac {\Omega}{{\cal M}^2}} ({\frac {\mu^2}{{\cal
M}^2}})^{a^2+\beta^2-1} K_{0c}\nonumber  \\ && - {\frac
{1}{2\pi^2}}a^2 ({\frac {\mu^2}{{\cal M}^2}})^{a^2} K_{0111}+ 2
a^4 ({\frac {\mu^2}{{\cal M}^2}})^{a^2}(K_{02})^2 \nonumber \\
&& -8\pi {\frac {\Omega}{{\cal M}^2}} a^2({\frac {\mu^2}{{\cal
M}^2}})^{a^2+\beta^2-1} K_{02} K_{0c}-{\frac {2}{\pi}}{\frac
{\Omega}{{\cal M}^2}} \beta^2({\frac {\mu^2}{{\cal M}^2}})^{a^2+\beta^2-1} K_{02c}\nonumber \\
&&+ {\frac {4}{\pi}}{\frac {\Omega}{{\cal M}^2}} a \beta ({\frac
{\mu^2}{{\cal M}^2}})^{a^2+\beta^2-1} K_{011s} + 8\pi^2 (\frac
{\Omega}{{\cal M}^2})^2 ({\frac {\mu^2}{{\cal
M}^2}})^{a^2+2\beta^2-2} (K_{0c})^2\nonumber \\ &&+ (\frac
{\Omega}{{\cal M}^2})^2 ({\frac {\mu^2}{{\cal
M}^2}})^{a^2+2\beta^2-2} K_{0ec11} + (\frac {\Omega}{{\cal
M}^2})^2 ({\frac {\mu^2}{{\cal M}^2}})^{a^2+2\beta^2-2} K_{0ee11}
\;,
\end{eqnarray}
where,
$$K_{02}\equiv \int^\infty_0 dx x K_0^2(x),  \hspace*{1.5cm}
\  \  \  \  \ \  \ K_{0c}\equiv \int^\infty_0 dx x [\cosh(4 a
\beta K_0 (x))-1]\;,$$
\begin{eqnarray*}
K_{0111} &=&\int^{2\pi}_0 d\theta_1\int^{2\pi}_0
d\theta_2\int^{\infty}_0 d\rho_1 \int^{\infty}_0 d\rho_2
\rho_1\rho_2 K_0(R)K_0(\rho_1)K_0(\rho_2)  \; ,
\\ K_{02c} &=&\int^{2\pi}_0 d\theta_1\int^{2\pi}_0
d\theta_2\int^{\infty}_0 d\rho_1 \int^{\infty}_0 d\rho_2
\rho_1\rho_2 (K_0(R))^2 [\cosh(4 a \beta K_0(\rho_2))-1] \; ,
\end{eqnarray*}
\begin{eqnarray*}
K_{011s} &=&\int^{2\pi}_0 d\theta_1\int^{2\pi}_0
d\theta_2\int^{\infty}_0 d\rho_1 \int^{\infty}_0 d\rho_2
\rho_1\rho_2 K_0(R)K_0(\rho_1)\sinh(4 a \beta K_0(\rho_2)) \; , \\
K_{0ec11} &=&\int^{2\pi}_0 d\theta_1\int^{2\pi}_0
d\theta_2\int^{\infty}_0 d\rho_1 \int^{\infty}_0 d\rho_2
\rho_1\rho_2 [\exp\{-4\beta^2K_0(R)\}-1]\nonumber \\ &&
\hspace*{5cm}[\cosh(4 a
\beta (K_0(\rho_1)+K_0(\rho_2)))-1]  \; , \\
K_{0ee11}&=&\int^{2\pi}_0 d\theta_1\int^{2\pi}_0
d\theta_2\int^{\infty}_0 d\rho_1 \int^{\infty}_0 d\rho_2
\rho_1\rho_2 [\exp\{4\beta^2K_0(R)\}-1]\nonumber \\ &&
\hspace*{5cm}[\exp\{-4 a \beta (K_0(\rho_1)-K_0(\rho_2))\}-1]
\end{eqnarray*}
with
$R=\sqrt{\rho_1^2-2\rho_1\rho_2\cos(\theta_1-\theta_2)+\rho_2^2}$.

In the right hand side of Eq.(12), the first two terms are the
expression of $G_a$ approximated up to the first order of
$\epsilon$, $G_a^{I}=[G_a^{(0)}+ \epsilon
G_a^{(1)}]_{\epsilon=1}$. To determine the arbitrary parameter
$\mu$ for $G_a^{I}$, as stated in the above, we can require that
${\frac {d G_a^{I}}{d(\mu^2)}}=0$ according to the PMS, and have
\begin{equation}
{\frac {\mu^2}{{\cal M}^2}}=\biggl(4 \pi {\frac {\Omega}{{\cal
M}^2}} K_{0c} {\frac {1-a^2-\beta^2}{a^2(1-2 a^2
K_{02})}}\biggl)^{1/(1-\beta^2)} \;.
\end{equation}
Thus, substituting last equation into the expression of $G_a^{I}$
gives approximately the result of $G_a$ up to the first order in
VPT. Note that ${\cal M}$ can be taken as any positive value, and
usually it can be referred to as unit when one renders various
quantities dimensionless for numerical calculations.

As for $\mu$ at the second order, following the next section, one
can numerically check that the condition ${\frac {d
G_a^{II}}{d(\mu^2)}}=0$ does not produce a real $\mu^2$ for a real
value of $a$. However, for a real $a$, $G_a$ should be real
because EVE's of odd-power fields in the sG field theory vanish
(See Eq.(2)). Hence we have to resort to ${\frac {d^2
G_a^{II}}{d(\mu^2)^2}}=0$ for determining $\mu$ at this order
(note that $\mu$ enters the scheme in its squared power $\mu^2$).
Fortunately, $\mu$ can explicitly be obtained with a long
expression from this condition, and then the approximate $G_a$ up
to the second order in VPT can be concretely given from Eq.(12).

In the same way, one can approximately give $G_a$ up to higher
orders. Here we do not continue to consider it, and next section
we will carry out a numerical calculation on the first and the
second order results.

In passing, $G_a^{(n)}$ in Eq.(8) is only the sum of all
$n$th-order connected Feynman diagrams consisting of the external
vertices from $\exp\{i \sqrt{8\pi}a \phi(0)\}$ and $n$ vertices
from ${\cal H}_I(\phi_{\vec{r}},\mu)$. Hence, one can also obtain
the above results Eq.(12) by borrowing Feynman diagrammatic
technique with the propagator of Eq.(6).

\section{Numerical Calculations and Comparisons}
\label{3}

To perform numerical calculations, we take ${\cal M}=1$ and all
physical quantities in Eq.(12) and (13) are dimensionless.
Simultaneously, one can check that this treatment (taking ${\cal
M}=1$) amounts to taking the same normalization conditions as
Eqs.(6) and (16) in Ref.~\cite{1}. Thus, our results can be
compared with the exact formula. From Eq.(12) in Ref.~\cite{1},
the dimensionless $\Omega$ in Eqs.(12) and (13) can be written as
\begin{equation}
\Omega={\frac {\Gamma(\beta^2)}{\pi
\Gamma(1-\beta^2)}}\biggl[{\frac {\sqrt{\pi}\Gamma({\frac
{1}{2}}+{\frac {\xi}{2}})}{2 \Gamma({\frac
{\xi}{2}})}}\biggl]^{2-2\beta^2}
\end{equation}
with $\xi={\frac {\beta^2}{1-\beta^2}}$. In last equation, we took
the soliton mass as unit to make $\Omega$ dimensionless. For
comparisons, we also do so for the conjectured exact formula
Eq.(20) in Ref.~\cite{1},$G_a^{exact}$, when it is employed.

Now we first report numerical results up to the first order.
Mathematica5.0 programm gives that $K_{02}=0.5$ and $K_{0c}$ is
finite for the case of $a \beta\le 0.426925$ which is involved in
the range of $a \beta<{\frac {1}{2}}$ for the conjectured exact
formula. Taking $\beta=0.2$ and $0.5$ as examples, we compared our
results $G_a^{I}$ with the conjectured exact results of
Ref.~\cite{1} in Figs.1 and 2, respectively.
\begin{figure}[h]
\includegraphics{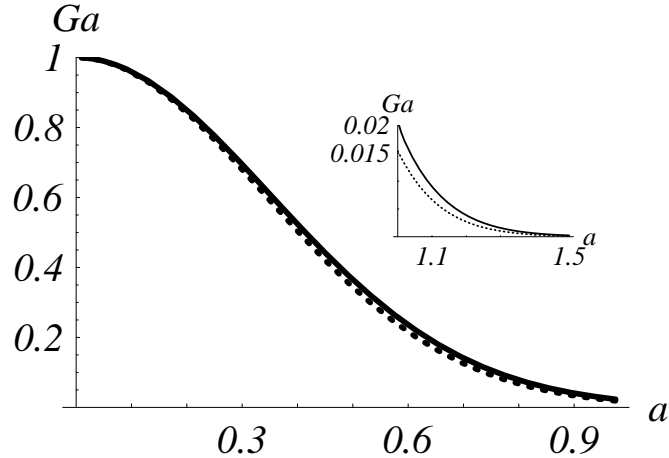}
\caption{\label{fig:1} Comparison between the first-order
($G_a^I$, solid curve) and conjectured exact ($G_a^{exact}$,
dashed curve) sG expectation values of exponential fields at
$\beta=0.2$.}
\end{figure}
\begin{figure}[h]
\includegraphics{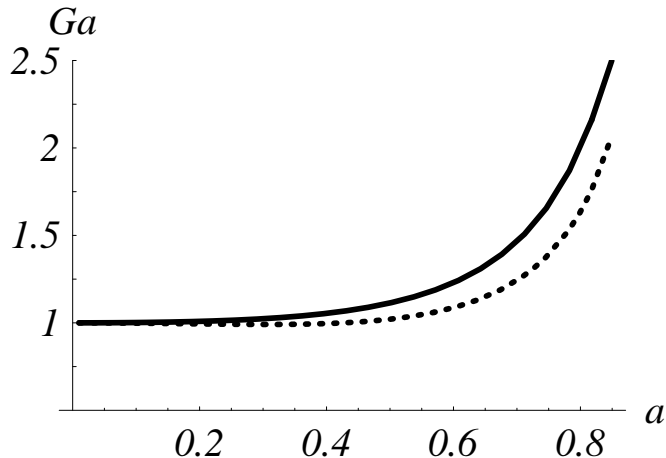}
\caption{\label{fig:2} Similar to Fig.1 but at $\beta=0.5$.}
\end{figure}
In Fig.1 and Fig.2, the solid curves are $G_a^I$, and the dashed
curves are $G_a^{exact}$. In Fig.1, $G_a$ decreases from $1$ to
$0$ with the increase of $a$. This tendency exists in the case of
$\beta<0.4$ or so. When $a<1$ and $a^2+\beta^2>1$, Eq.(13) will
produce a complex $\mu^2$ and one should use the second derivative
for determining $\mu$. For sufficiently small $\beta$, one needn't
do so because $G_a$ tends to zero with the increase of $a$ before
$\mu^2$ approaches complex values. In Fig.2, $G_a$ increases
almost from $1$ with the increase of $a$. This tendency appears
for other larger values of $\beta$. These figures indicate that
for the case of $a<0.2$ or so, the first-order results almost
completely agree with the conjectured exact results, and for
larger $a$, our results differ from the exact results with about
ten percents or so (at most with 20 more percents when $a$
approaches values with $a \beta= 0.426925$ satisfied, see Table
I).

For calculating $G_a^{II}$, we need to calculate integrals in
Eq.(12) which are involved in the zeroth order modified Bessel
functions of the second kind. Noting Gegenbauer's addition formula
for the zeroth-order modified Bessel functions of the second kind
\cite{15} $K_0(R)=I_0(\rho_1)K_0(\rho_2)+2\sum_{n=1}^\infty
\cos[n(\theta_1-\theta_2)]I_n(\rho_1)K_n(\rho_2)$ with
$\rho_1<\rho_2$ ($I_n(\rho_1)$ is the $n$th-order modified Bessel
functions of the first kind), one can finish those integrals by
dividing the plane \{$\rho_1$,$\rho_2$\} into two parts: one part
with $\rho_1<\rho_2$ and the other with $\rho_1>\rho_2$. In
performing the calculations, because of the oscillatory property
of $\cos[n(\theta_1-\theta_2)]$, it is enough to truncate the
series in the Gegenbauer's addition formula at some $n$. Finishing
those integrals for $\beta=0.5$ with long-time calculations of
Mathematica5.0 system, we obtained the results of $G_a^{II}$ and
accordingly compared them with $G_a^I$ and the conjectured exact
results in the table.
\begin{table}[h]
\caption{Comparisons: $G_a^{II}$, $G_a^I$ and $G_a^{exact}$ for
$\beta=0.5$}
\begin{tabular}{|l|l|l|l|l|l|}
 \hline\hline
$a$&$G_a^{exact}$&$G_a^I$&$G_a^{II}$&$\Delta^{I}$&$\Delta^{II}$
\\ \hline 0.01&0.999981 &1.00001758&1.00001187
&0.003661&0.003090  \\ \hline 0.1&0.998193
&1.001847531 &1.00137859 &0.36542 &0.3185 \\
\hline 0.2&0.993954& 1.00850388 &1.00678750 &1.4639&1.2912
\\ \hline 0.3&0.990960 &1.02358276
&1.01961512 &3.2920&2.8917
\\  \hline 0.4&0.995987 &1.05423835
&1.04769147&5.8486&5.1913
\\  \hline 0.5&1.020682&1.11390679
&1.10415979 &9.1336&8.1786
\\  \hline 0.6&1.086765&1.22971602
&1.21627955 &13.1538&11.9174
\\  \hline  0.7&1.242866
&1.46517988 &1.44784716 &17.8872&16.4926
\\  \hline 0.8&1.634342&2.00162788
&1.99288037 &22.4730&21.9378
\\  \hline\hline
\end{tabular}
\end{table}
In Table I, $\Delta^I={\frac
{G_a^I-G_a^{exact}}{G_a^{exact}}}\times 100$ and
$\Delta^{II}={\frac {G_a^{II}-G_a^{exact}}{G_a^{exact}}}\times
100$. This table indicates that $G_a^{II}$ has an about one
percent smaller difference from the conjectured exact result than
$G_a^{I}$ at $\beta=0.5$. We also checked the cases of smaller and
larger $\beta$ ($\beta^2<0.426925$), and found the same
conclusion.

\section{Conclusion}
\label{5}

In this letter, in order to check the conjectured exact formula in
Ref.~\cite{1}, we have calculated the sG expectation Value of the
exponential field $e^{ia\phi(\vec{r})}$ up to the second order
with variational perturbation approach. According to our numerical
results in last section, for not large $a$, the exact formula
conjectured by Lukyanov and Zamolodchikov in Ref.~\cite{1} is
correct for all values of $\beta$ in the range of $\beta^2<1$, and
for larger $a$ ($|Re(a)|<1/(2\beta)$), our numerical results can
be believed to support the conjectured exact formula since our
method is an approximate one. Thus, from the existed reports in
Refs.~\cite{1,4,5} and our report here, we believe that the
conjectured exact formula in Ref.~\cite{1} is completely
convincible.

As ending the present letter, we mention some interesting
problems. The normal-ordering prescription amounts to a
renormalization procedure on the mass parameter and makes $G_a^I$
finite for the range of $a \beta<{\frac {1}{2}}$. If introducing
additional renormalization scheme on the coupling $\beta$ and the
exponential-field parameter $a$, it will be possible to obtain a
finite $G_a$ for the range of $a \beta>{\frac {1}{2}}$.
Furthermore, it will be also interesting to calculate the sG
expectation values of the exponential fields on the asymmetrical
vacua. On the other hand, because it was an important progress in
calculating VEV's of local fields, the Lukyanov-Zamolodchikov
conjecture in Ref.~\cite{1} has stimulated generalizations or
conjectures on the VEV's of local fields in some field theories
\cite{16}, and hence the method in the present letter and its
generalization to finite temperature case can be used to check or
confirm those generalizations or conjectures in Ref.~\cite{4,16}.
Besides, although our results up to the second order improved the
results up to the first order, Table I indicates that the
improvement is small. Since the Lukyanov-Zamolodchikov conjecture
can be believed to be correct, comparing it with higher-order
contributions in VPT for the sG expectation Values of the
exponential fields will provide the first check in QFT on the
convergency of VPT.

\begin{acknowledgments}
I acknowledges Prof. C. F. Qiao for his useful discussions on VPT
and helps in numerical calculations. This project was sponsored by
SRF for ROCS, SEM and supported by the National Natural Science
Foundation of China.
\end{acknowledgments}


\begin{thebibliography}{99}
\bibitem{1} S. Lukyanov and A. Zamolodchikov, Nucl. Phys. B {\bf 493} (1997) 571.
\bibitem{2} S. Coleman, Phys. Rev. D {\bf 11} (1975) 2088.
\bibitem{3} A. Zamolodchikov and Al. Zamolodchikov, Nucl. Phys. B {\bf 477} (1996) 577.
\bibitem{4} V. Fateev, S. Lukyanov, A. Zamolodchikov and Al. Zamolodchikov, Phys. Lett. B {\bf 406} (1997) 83;
            Nucl. Phys. B {\bf 516} (1998) 652;
\bibitem{5} R. H. Poghossian, Nucl. Phys. B {\bf 570} (2000) 506; V. V. Mkhitaryan, R. H. Poghossian, and T. A.
            Sedrakyan, J. Phys. A {\bf 33} (2000) 3335.
\bibitem{5a}Z. Bajnok, L. Palla, G. Tak$\acute{a}$cs and F.
            W$\acute{a}$gner, Nucl. Phys. B {\bf 587} (2000) 585 .
\bibitem{6} H. Kleinert and V. Schulte-Frohlinde, {\it Critical
            Properties of $\phi^4$-Theories}, World Scientific, Singapore, 2001, Chapter 19;
             H. Kleinert, {\it   Path Integrals in Quantum Mechanics, Statistics, Polymer Physics, and Financial Markets
             }, 3rd Ed., World Scientific, Singapore, 2004.
\bibitem{7} P. M. Stevenson, Phys. Rev. D {\bf 23} (1981) 2916;
            Phys. Lett. B {\bf 100} (1981) 61; Phys. Rev. D {\bf 24} (1981)
            1622; S. K. Kaufmann and S. M. Perez, J. Phys. A {\bf 17} (1984)
            2027; P. M. Stevenson, Nucl. Phys. B {\bf 231} (1984) 65.
\bibitem{8} L. Cohen, Proc. Phys. Soc. A 68 (1955) 419;425.
\bibitem{9} W. F. Lu, C. K. Kim and K. Nham, Phys. Lett. B {\bf 546} (2002) 177.
\bibitem{10}A. Okopinska, Phys. Rev. D {\bf 35} (1987) 1835 .
\bibitem{11}I. Stancu and P. M. Stenvenson,Phys. Rev. D {\bf 42} (1990) 2710.
\bibitem{12}S. J. Chang, Phys. Rev. D {\bf 13} (1976) 2778.
\bibitem{13}W. F. Lu and C. K. Kim, J. Phys. A {\bf 35} (2002) 393.
\bibitem{14}I. S. Gradshteyn and I. M. Ryzhik, {\it Table of Integrals, Series, and Products}, 4th
            Edition, Academic Press, New York, 1980 .
\bibitem{15}A. Gray and G. B. Mathews, {\it A Treatise Bessel Functions and Their Applications to
            Physics}, Dover Publications, Inc., New York, 1966, pp.74 .
\bibitem{16}P. Baseilhac and V. A. Fateev, Nucl. Phys. {\bf 532} (1998) 567;
             C. Ahn, P. Baseilhac, V. A. Fateev, C. Kim and C. Rim, Phys. Lett. B {\bf 481} (2000) 114;
            C. Ahn, P. Baseilhac, C. Kim and C. Rim, Phys. Rev. D {\bf 64} (2001) 046002;
            S. Lukyanov, Nucl. Phys. B {\bf 612} (2001) 391.
\end{thebibliography}
\end{document}